\documentclass[lettersize,hidelinks,conference]{IEEEtran}

\usepackage{amsmath,amssymb,amsfonts}
\usepackage{array}
\usepackage{subfigure}
\usepackage{textcomp}
\usepackage{stfloats}
\usepackage{url}
\usepackage{verbatim}
\usepackage{graphicx}
\usepackage{cite}
\usepackage{xcolor}
\usepackage{orcidlink}
\usepackage[acronym,shortcuts]{glossaries}
\usepackage{bm}
\usepackage{relsize}
\usepackage{soul}
\usepackage{algorithm, algpseudocode}
\usepackage{algcompatible}
\usepackage{mathtools}
\usepackage{bm}
\usepackage{lipsum}
\usepackage{mathtools}
\usepackage{lipsum}
\usepackage[hang,flushmargin]{footmisc}
\usepackage{tabularx}

\def\BibTeX{{\rm B\kern-.05em{\sc i\kern-.025em b}\kern-.08em
T\kern-.1667em\lower.7ex\hbox{E}\kern-.125emX}}

\newcommand{\trans}[0]{^{\mathsf{T}}}

\newcommand{\herm}[0]{^{\mathsf{H}}}

\newcommand{\Real}[1]{\Re\{{#1}\}}
\newcommand{\Imag}[1]{\Im\{{#1}\}}

\newacronym{PCA}{PCA}{principal component analysis}
\newacronym{MDSM}{MDSM}{multi-domain sparse modulation}
\newacronym{P2P}{P2P}{point-to-point}
\newacronym{OTAC}{AirComp}{over-the-air computation}
\newacronym{TX}{TX}{transmitter}
\newacronym{RX}{RX}{receiver}
\newacronym{IoT}{IoT}{Internet of Things}
\newacronym{AI/ML}{AI/ML}{artifitial intelligence/machine learning}
\newacronym{SDR}{SDR}{semi-definite relaxation}
\newacronym{EVD}{EVD}{eigenvalue decomposition}
\newacronym{GR}{GR}{Gaussian randomization}
\newacronym{SCA}{SCA}{successive convex approximation}
\newacronym{BnB}{BnB}{branch and bound}
\newacronym{QT}{QT}{quadratic transform}
\newacronym{RQ}{RQ}{Rayleigh quotient}
\newacronym{SOCP}{SOCP}{second-order cone programming}
\newacronym{CDF}{CDF}{cumulative distribution function}
\newacronym{UF}{UF}{uniform-forcing}
\newacronym{AP}{AP}{access point}
\newacronym{RSDR}{R-SDR}{regularized semi-definite relaxation}
\newacronym{R-SDR}{R-SDR}{regularized SDR}
\newacronym{flops}{flops}{floating point operations}
\newacronym{ED}{ED}{edge device}
\newacronym{SINR}{SINR}{signal to interference-plus-noise ratio}
\newacronym{SIC}{SIC}{successive interference cancellation}
\newacronym{CSI}{CSI}{channel state information}
\newacronym{LoS}{LoS}{line-of-sight}
\newacronym{NLoS}{NLoS}{non-LoS}
\newacronym{RPE}{RPE}{radar parameter estimation}
\newacronym{OTFS}{OTFS}{orthogonal time frequency space}
\newacronym{AFDM}{AFDM}{affine frequency division multiplexing}
\newacronym{CRLB}{CRLB}{Cram{\`e}r-Rao lower bound}
\newacronym{BCRLB}{BCRLB}{Bayesian Cram{\`e}r-Rao lower bound}
\newacronym{BBI}{BBI}{Bayesian bilinear inference}
\newacronym{AoA}{AoA}{angle-of-arrival}
\newacronym{SNR}{SNR}{signal-to-noise ratio}
\newacronym{ML}{ML}{maximum likelihood}
\newacronym{MIMO}{MIMO}{multiple-input multiple-output}
\newacronym{MISO}{MISO}{multiple-input single-output}
\newacronym{SIMO}{SIMO}{single-input multiple-output}
\newacronym{SISO}{SISO}{single-input single-output}
\newacronym{MUSIC}{MUSIC}{multiple signal classification}
\newacronym{MU}{MU}{multi-user}
\newacronym{ROOT-MUSIC}{ROOT-MUSIC}{ROOT multiple signal classification}
\newacronym{JCAS}{JCAS}{joint communication and sensing}
\newacronym{JCR}{JCR}{joint communications and radar}
\newacronym{ISAC}{ISAC}{integrated sensing and communications}
\newacronym{3D}{3D}{three-dimensional}
\newacronym{2D}{2D}{two-dimensional}
\newacronym{1D}{1D}{one-dimensional}
\newacronym{BF}{BF}{beamforming}
\newacronym{ROI}{ROI}{region of interest}
\newacronym{mmWave}{mmWave}{millimeter-wave}
\newacronym{MF}{MF}{matched-filter}
\newacronym{DD}{DD}{delay-Doppler}
\newacronym{SotA}{SotA}{state-of-the-art}
\newacronym{ULA}{ULA}{uniform linear array}
\newacronym{QAM}{QAM}{quadrature amplitude modulation}
\newacronym{ISFFT}{ISFFT}{inverse symplectic finite Fourier transform}
\newacronym{SFFT}{SFFT}{symplectic finite Fourier transform}
\newacronym{ISI}{ISI}{inter-symbol interference}
\newacronym{AWGN}{AWGN}{additive white Gaussian noise}
\newacronym{MSE}{MSE}{mean-squared-error}
\newacronym{LMMSE}{LMMSE}{linear minimum mean square error}
\newacronym{RMSE}{RMSE}{root mean square error}
\newacronym{ESPRIT}{ESPRIT}{estimation of signal parameters via rotational invariant techniques}
\newacronym{OFDM}{OFDM}{orthogonal frequency division multiplexing}
\newacronym{OCDM}{OCDM}{orthogonal chirp division multiplexing}
\newacronym{BS}{BS}{base station}
\newacronym{UE}{UE}{user equipment}
\newacronym{JCEDD}{JCEDD}{joint channel estimation and data detection}
\newacronym{PDA}{PDA}{probabilistic data association}
\newacronym{PMF}{PMF}{probability mass function}
\newacronym{PBiGaBP}{PBiGaBP}{parametric bilinear Gaussian belief propagation}
\newacronym{PBiGAMP}{PBiGAMP}{parametric bilinear generalized approximate message passing}
\newacronym{GaBP}{GaBP}{Gaussian belief propagation}
\newacronym{FT}{FT}{frequency-time}
\newacronym{DFT}{DFT}{discrete Fourier transform}
\newacronym{IDFT}{IDFT}{inverse discrete Fourier transform}
\newacronym{TD}{TD}{time domain}
\newacronym{wlg}{w.l.g.}{without loss of generality}
\newacronym{CP}{CP}{cyclic prefix}
\newacronym{DAF}{DAF}{discrete affine Fourier}
\newacronym{DAFT}{DAFT}{discrete affine Fourier transform}
\newacronym{IDAFT}{IDAFT}{inverse discrete affine Fourier transform}
\newacronym{CPP}{CPP}{\textit{chirp-periodic} prefix}
\newacronym{IDZT}{IDZT}{inverse discrete Zak transform}
\newacronym{DZT}{DZT}{discrete Zak transform}
\newacronym{P/S}{P/S}{parallel-to-serial}
\newacronym{S/P}{S/P}{serial-to-parallel}
\newacronym{SBL}{SBL}{sparse Bayesian learning}
\newacronym{MPA}{MPA}{message passing algorithms}
\newacronym{EM}{EM}{expectation maximization}
\newacronym{sIC}{soft IC}{soft interference cancellation}
\newacronym{soft RG}{soft RG}{soft replica generation}
\newacronym{BG}{BG}{belief generation}
\newacronym{SGA}{SGA}{scalar Gaussian approximation}
\newacronym{CLT}{CLT}{central limit theorem}
\newacronym{PDF}{PDF}{probability density function}
\newacronym{QPSK}{QPSK}{quadrature phase-shift keying}
\newacronym{ICI}{ICI}{inter-carrier interference}
\newacronym{BER}{BER}{bit error rate}
\newacronym{DoF}{DoF}{degrees-of-freedom}
\newacronym{VGA}{VGA}{vector Gaussian approximation}
\newacronym{FD}{FD}{full-duplex}
\newacronym{NMSE}{NMSE}{normalized mean square error}
\newacronym{KL}{KL}{Kullback-Leibler}
\newacronym{LASSO}{LASSO}{least absolute shrinkage and selection operator}
\newacronym{FP}{FP}{fractional programming}
\newacronym{CC}{CC}{communication-centric}
\newacronym{RC}{RC}{raised-cosine}
\newacronym{RRC}{RRC}{root raised-cosine}
\newacronym{6G}{6G}{sixth-generation}
\newacronym{V2X}{V2X}{vehicle-to-everything}
\newacronym{LEO}{LEO}{low-earth orbit}
\newacronym{I/O}{I/O}{input-output}
\newacronym{CE}{CE}{channel estimation}
\newacronym{ICC}{ICC}{integrated communication and computing}
\newacronym{ISCC}{ISCC}{integrated sensing, communications and computing}
\newacronym{PAM}{PAM}{pulse amplitude modulation}
\newacronym{iid}{i.i.d.}{independent and identically distributed}
\newacronym{MEC}{MEC}{mobile edge computing}
\newacronym{REMS}{REMS}{reconfigurable electromagnetic structure}
\newacronym{RIS}{RIS}{reconfigurable intelligent surface}
\newacronym{MMSE}{MMSE}{minimum mean square error}
\newacronym{DPC}{DPC}{dirty paper coding}
\newacronym{SER}{SER}{Symbol Error Rate}

\begin{document}


\title{Computing on Dirty Paper: Interference-Free Integrated Communication and Computing}

\author{\IEEEauthorblockN{Kuranage Roche Rayan Ranasinghe,$\!\!^*$ Giuseppe Thadeu Freitas de Abreu,$\!\!^*$ David~Gonz{\'a}lez~G.$\!^\dag$ and Carlo Fischione$^\ddag$}
\IEEEauthorblockA{$^*$\textit{School of Computer Science and Engineering, Constructor University, Bremen, Germany} \\
$^\dag$\textit{Wireless Communications Technologies Group, AUMOVIO SE, Frankfurt am Main, Germany} \\
$^\ddag$\textit{School of Electrical Engineering and Computer Science, KTH Royal Institute of Technology, Stockholm, Sweden} \\
Emails: [kranasinghe, gabreu]@constructor.university, david.gonzalez.g@ieee.org, carlofi@kth.se}}


\maketitle

\begin{abstract}

Inspired by Costa's pioneering work on \ac{DPC}, this paper proposes a novel scheme for \ac{ICC}, named \textit{Computing on Dirty Paper}, whereby the transmission of discrete data symbols for communication, and \ac{OTAC} of nomographic functions can be achieved simultaneously over common multiple-access channels.
In particular, the proposed scheme allows for the integration of communication and computation in a manner that is asymptotically interference-free, by pre-canceling the computing symbols at the \acp{TX} using \ac{DPC} principles.
A simulation-based assessment of the proposed \ac{ICC} scheme under a \ac{SIMO} setup is also offered, including the evaluation of  performance for data detection, and of \ac{MSE} performance for function computation, over a block of symbols.
The results validate the proposed method and demonstrate its ability to significantly outperform \ac{SotA} \ac{ICC} schemes in terms of both \ac{BER} and \ac{MSE}.

\end{abstract}


\glsresetall

\IEEEpeerreviewmaketitle

\vspace{-1ex}
\section{Introduction}
\label{sec:introduction}
\vspace{-1ex}

The rapid proliferation of \ac{IoT} devices and edge computing applications is driving the need for seamless integration of communication and computation in wireless networks, forming the basis for the \ac{ICC} paradigm \cite{WenCOMST2024,QiIEEENET2024}.
This approach aims to reduce latency and resource overhead by enabling simultaneous data transmission and the \ac{OTAC} \cite{NazerTIT07, LiuTWC20, QiTWC2021,AndoIOTJ2025} of nomographic functions over shared channels, particularly in multiple-access scenarios where interference poses significant challenges \cite{HuangTWC2025}. 

Traditional methods treat the communication and computation operations separately, leading to inefficiencies, but recent advancements in \ac{ICC} have shown that joint designs can significantly enhance performance \cite{ranasinghe2025flexibledesignframeworkintegrated}.
However, when discrete data symbols must be communicated alongside computing signals, mutual interference between the two tasks degrades performance, necessitating innovative interference management techniques \cite{YeSPAWC2024}. 

\Ac{DPC}, a foundational concept in information theory first proposed in \cite{CostaTIT1983}, addresses similar issues by allowing a transmitter to communicate reliably in the presence of known interference through precoding, effectively ``writing on dirty paper'' to achieve the same channel capacity as if the interference were absent. 

While many ways to implement \ac{DPC} have been proposed including approaches leveraging Tomlinson-Harashima precoding \cite{ZhangWCNC2023}, nested lattice codes have emerged as a particularly effective method due to their structured nature and ability to approach the theoretical limits of \ac{DPC} \cite{ZamirTIT2002,ErezTIT2005,LiuTIT2006,KochmanTIT2009,LinTCOMM2011}.
In addition, \ac{DPC} has also been successfully applied in various multifunctional communication system aspects, such as in the \ac{ISAC} paradigm \cite{LiICC2022,nikbakht2025integrated,nikbakht2025mimo}.

Inspired by these developments, we introduce in this paper \textit{Computing on Dirty Paper}, a novel \ac{ICC} approach that leverages \ac{DPC} to pre-cancel computing symbols at the users in a \ac{SIMO} uplink setting, enabling the simultaneous transmission of discrete data and the computation of nomographic functions over the same multiple access channel. 
In particular, we demonstrate that typical \ac{DPC} schemes such as the one proposed in 
\cite{ErezTIT2005} can be incorporated at the \ac{TX}, and combined with a classic \ac{LMMSE} estimator at the \ac{RX}, to enable \ac{ICC} with no impact of the \ac{OTAC} operation over the \ac{BER} of the communication stream.
A simulation-based assessment of the method in a \ac{SIMO} setup is performed, which validates our approach, showing substantial improvements over \ac{SotA} benchmarks in terms of both functionalities.

\textit{Notation:} Throughout the manuscript, vectors and matrices are represented by lowercase and uppercase boldface letters, respectively;
$\mathbf{I}_M$ denotes an identity matrix of size $M$ and $\mathbf{1}_M$ denotes a column vector composed of $M$ ones; 
the Euclidean norm and the absolute value of a scalar are respectively given by $\|\cdot\|_2$ and $|\cdot|$;
the transpose and hermitian operations follow the conventional form $(\cdot)\trans$ and $(\cdot)\herm$, respectively;
$\Re{\{\cdot\}}$, $\Im{\{\cdot\}}$ and  $\mathrm{min}(\cdot)$ represents the real part, imaginary part and the minimum operator, respectively.
Finally, $\sim \mathcal{CN}(\mu,\sigma^2)$ denotes the complex Gaussian distribution with mean $\mu$ and variance $\sigma^2$, where $\sim$ denotes ``is distributed as''.

\vspace{-1ex}
\section{System and Signal Models}
\label{sec:system_model}
\vspace{-1ex}

\begin{figure}[H]
\centering
\includegraphics[width=1\columnwidth]{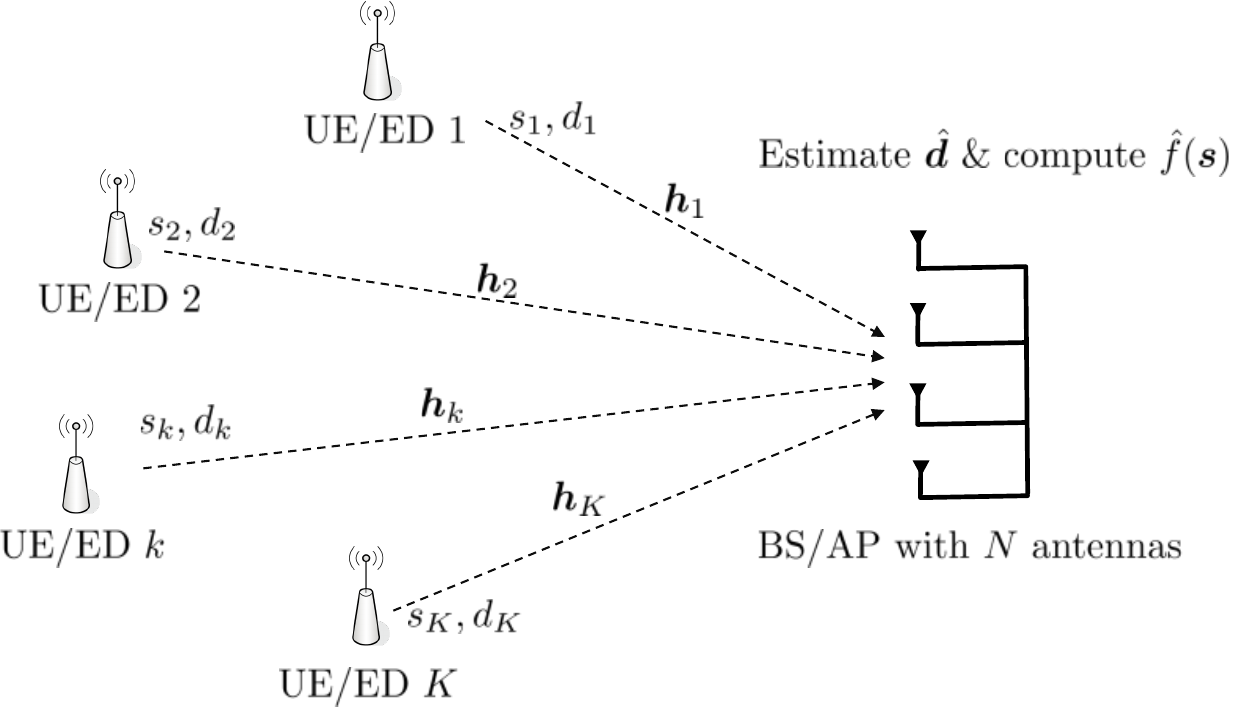}
\vspace{-4ex}
\caption{\ac{SIMO} \ac{ICC} system consisting of a \ac{RX} with $N$ antennas and $K$ single antenna users.}
\label{fig:system_model}
\end{figure}

Consider a typical \ac{SIMO} uplink setup composed of $K$ single-antenna \ac{UE}/\acp{ED} and one \ac{BS}/\ac{AP} equipped with $N$ antennas, as illustrated in Fig. \ref{fig:system_model}.
Throughout this article we refer to the \acp{UE}/\acp{ED} as users, unless otherwise specified, and refer to the \ac{BS}/\ac{AP} as the \ac{RX}.

\vspace{-3ex}
\subsection{Uplink ICC Signal Model}

Under the assumption of perfect symbol synchronization amongst users, the received signal $\bm{y}^{(t)} \in \mathbb{C}^{N\times 1}$ at the \ac{BS}/\ac{AP} subjected to fading and noise at a $t$-th time slot is given by
\vspace{-1ex}
\begin{equation}
\label{eq:received_signal}
\vspace{-1ex}
\bm{y}^{(t)} = \sum_{k=1}^K \bm{h}_k^{(t)}  {x}_k^{(t)} + \bm{w}^{(t)},
\vspace{-0.5ex}
\end{equation}
where ${x}_k^{(t)} \in \mathbb{C}$ is a multifunctional transmit signal from user $k$, incorporating communication and computing signals; $\bm{w}^{(t)} \in \mathbb{C}^{N\times1}\sim \mathcal{CN}(0,\sigma^2_w\mathbf{I}_N)$ is the \ac{AWGN} vector, and $\bm{h}_k^{(t)} \in \mathbb{C}^{N \times 1}$ is the \ac{SIMO} channel vector of user $k$ to the \ac{RX} following the uncorrelated block Rayleigh fading model typically assumed in the \ac{OTAC} literature \cite{LiuTWC20,AndoCAMSAP2023}, such that the $(n,k)$-th elements $h_{n,k}\sim \mathcal{CN}(0,1)$ of the channel matrix $\bm{H}^{(t)}$ are assumed to be \ac{iid} circularly symmetric complex Gaussian random variable with zero mean and unit variance, and sufficient coherence time.

\vspace{-1ex}
\subsection{SotA Transmit Signal Design}
\label{subsec:transmission_scheme}

Under the \ac{SotA} \ac{ICC} paradigm \cite{RanasingheICNC_comp_2025,ranasinghe2025flexibledesignframeworkintegrated}, a trivial way to model the transmit signal is to decompose it as a sum of communication and computing components, $i.e.,$
\vspace{-0.5ex}
\begin{equation}
\label{eq:transmit_sig_decomposition}
x_k^{(t)} \triangleq d_k^{(t)} +  \psi_k(s_k^{(t)}),
\end{equation}
where $d_k^{(t)} \in \mathcal{D}$ and $s_k^{(t)} \in \mathcal{S}$ denote the modulated symbol for communication and computing for user $k$, respectively, with $\mathcal{D}$ representing an arbitrarily scaled discrete constellation of cardinality $D$, $e.g.$ \ac{QAM}; $\mathcal{S}$ representing a discrete set following \cite{RazavikiaTCOMM2024} for the computing symbols; while $\psi_k(\cdot)$ denotes a general pre-processing function for \ac{OTAC} to be elaborated in the following subsection.

The \ac{SotA} \ac{RX} design \cite{ranasinghe2025flexibledesignframeworkintegrated} typically consists of a two-stage approach, where the \ac{RX} first estimates the individual communication symbols $\{d_k^{(t)}\}_{k=1}^K$ per time slot via a typical \ac{LMMSE} or a lower complexity \ac{GaBP} technique under the assumption that the computing signal is noise, followed by the evaluation of a target function $f(\bm{s}^{(t)})$ via a combiner.

For future convenience, the received signal can now be reformulated in terms of matrices as
\begin{equation}
\bm{y}^{(t)} =  \bm{H}^{(t)}  \bm{x}^{(t)} + \bm{w}^{(t)} =\bm{H}^{(t)}  \big(\bm{d}^{(t)} + \bm{s}^{(t)}\big) + \bm{w}^{(t)},
\label{eq:received_signal_matrix}
\end{equation}
where the complex channel matrix $\bm{H}^{(t)} \triangleq [\bm{h}_1^{(t)},\dots, \bm{h}_K^{(t)}] \in \mathbb{C}^{N\times K}$, the concatenated transmit signal $\bm{x}^{(t)} \triangleq [x_1^{(t)},\dots, x_K^{(t)}]\trans \in \mathbb{C}^{K\times1}$, the data signal vector $\bm{d}^{(t)} \triangleq [d_1^{(t)},\dots, d_K^{(t)}]\trans \in \mathcal{D}^{K \times 1} \subset \mathbb{C}^{K\times1}$ and the computing signal vector $\bm{s}^{(t)} \triangleq [\psi_1(s_1^{(t)}),\dots, \psi_K(s_K^{(t)})]\trans \in \mathcal{S}^{K \times 1} \subset \mathbb{C}^{K\times1}$ are explicitly defined.

In addition, concatenating the received signals over $T$ time slots and assuming that the channel $\bm{H}^{(t)}$ (hereafter denoted as $\bm{H}$ for brevity) remains unchanged\footnote{The extension to time-varying channels is trivial under perfect \ac{CSI} at the \ac{RX}, and will be done in a follow-up work.}, we may express the received signal matrix as
\begin{equation}
\bm{Y} = \bm{H} \bm{X} + \bm{W} = \bm{H}  (\bm{D} + \bm{S}) + \bm{W},
\label{eq:received_signal_matrix_T}
\end{equation}
where $\bm{Y} \triangleq [\bm{y}^{(1)},\dots, \bm{y}^{(T)}] \in \mathbb{C}^{N\times T}$, $\bm{X} \triangleq [\bm{x}^{(1)},\dots, \bm{x}^{(T)}] \in \mathbb{C}^{K\times T}$, $\bm{D} \triangleq [\bm{d}^{(1)},\dots, \bm{d}^{(T)}] \in \mathcal{D}^{K\times T} \subset \mathbb{C}^{K\times T}$, $\bm{S} \triangleq [\bm{s}^{(1)},\dots, \bm{s}^{(T)}] \in \mathcal{S}^{K \times T} \subset \mathbb{C}^{K\times T}$ and $\bm{W} \triangleq [\bm{w}^{(1)},\dots, \bm{w}^{(T)}] \in \mathbb{C}^{N\times T}$ are the corresponding matrices.

Finally, for the sake of derivation of the proposed scheme, we assume that the computing symbols $\bm{s}^{(t)}$ across a block of $T$ time slots are identical (repeatedly transmitted), $i.e.,$ $s_k^{(t)} = s_k, \forall k$, and $\bm{s}^{(t)} = \bm{s}, \forall t$, and thus $\bm{S} = [\bm{s},\dots, \bm{s}] \in \mathcal{S}^{K \times T} \subset \mathbb{C}^{K\times T}$.

\vspace{-2ex}
\subsection{Description of the AirComp Operation}

The \ac{OTAC} operation consists of the evaluation of a target function $f(\bm{s})$ at the \ac{RX}, which can be described as \cite{LiuTWC20}
\vspace{-1ex}
\begin{equation}
\label{eq:target_function_def}
f(\bm{s}) = \phi\bigg(\sum_{k=1}^K \psi_k(s_k)\bigg),
\vspace{-1ex}
\end{equation}
where $\phi$ represents the \ac{OTAC} post-processing function for a general nomographic expression.

While many examples of nomographic functions are often considered in the \ac{OTAC} literature \cite{NazerTIT07,Wang_arxiv_2024}, we choose the arithmetic sum operaton for ease of exposition, given by
\vspace{-1ex}
\begin{equation}
\label{eq:target_function_def_SUM}
f(\bm{s}) = \phi\bigg(\sum_{k=1}^K \psi_k(s_k)\bigg) = \sum_{k=1}^K s_k,
\vspace{-1ex}
\end{equation}
where the corresponding pre- and post-processing functions are defined as $\psi_k(s_k) \triangleq s_k$ and $\phi\big(\sum_{k=1}^K \psi_k(s_k)\big) \triangleq \sum_{k=1}^K \psi_k(s_k)$.

\vspace{-1ex}
\section{Computing on Dirty Paper}
\label{sec:Computing_on_Dirty_Paper}

\subsection{Fundamentals on Dirty Paper Coding}
\label{subsec:DPC_fundamentals}

\Acf{DPC}, first introduced by Costa in \cite{CostaTIT1983}, is a powerful coding technique that allows the transmission of information over a noisy channel in the presence of known interference at the \acp{TX}, without being affected by the latter.
In more detail, the \ac{TX} -- the arbitrary \acp{UE} or \acp{ED} in our case -- needs to transmit a signal $d$ subject to interference $s$.
Then, the \ac{RX} receives signal $y = d + s + w$, where $w$ is random noise.
When $s$ is unknown to both the \ac{TX} and \ac{RX}, it can only be incorporated into the noise as previously proposed in \cite{ranasinghe2025flexibledesignframeworkintegrated} for the \ac{ICC} paradigm.
However, when $s$ is known to the \ac{TX}, the \ac{RX} can completely remove the interference $s$ by leveraging \ac{DPC}, even though it does not know $s$.
In this paper, we introduce a novel scheme for the \ac{DPC} operation in the context of \ac{ICC}, termed \textit{Computing on Dirty Paper}.

\subsection{Fundamentals on Nested Lattice Codes}
\label{subsec:nested_lattice_codes}

\begin{figure}[H]
\centering
\includegraphics[width=\columnwidth]{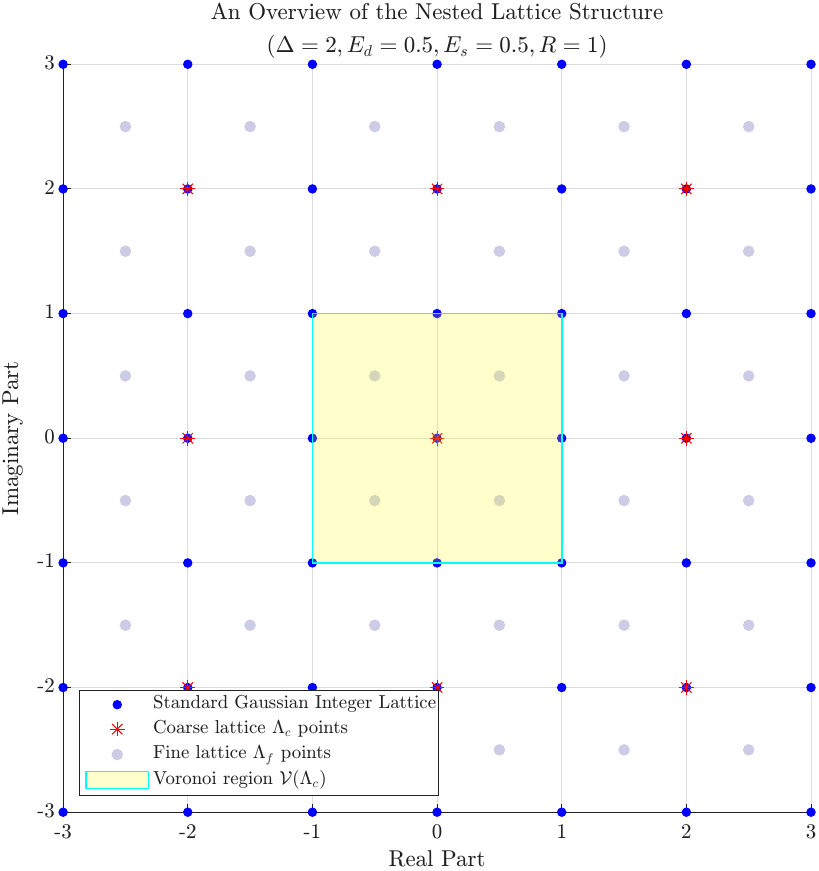}
\vspace{-3ex}
\caption{Nested lattice structure with coarse lattice $\Lambda_c$ and fine lattice $\Lambda_f$.}
\label{fig:lattice_DPC}
\vspace{-2ex}
\end{figure}

For the design of the lattice, consider a standard Gaussian integer lattice $\mathbb{Z}[j] = \{ a + bj \mid a, b \in \mathbb{Z} \}$ with generator matrix defined to be $\mathbf{I}_2$, as illustration in Fig. \ref{fig:lattice_DPC}, which is simple\footnote{A potential extension would be the use of Einstein integers for hexagonal packing.} and effective for complex signals (equivalent to a square lattice in $\mathbb{R}^2$) in a similar fashion to \cite{ErezTIT2005} as follows:

\textbf{Coarse lattice:} Let $\Lambda_c = \Delta \mathbb{Z}[j] \subset \mathbb{C}$, where $\Delta > 0$ sets the modulo region; $i.e.,$ the ``coarse lattice cube'' bounded by sides of length $[-\Delta/2, \Delta/2]$.
This choice ensures that the transmit power is bounded during the modulo operation.

\textbf{Fine lattice:} Let $\Lambda_f = \delta \mathbb{Z}[j]$, with $\delta = \Delta / \sqrt{M}$ and $M = 2^{2R}$ where $M$ represents the (integer) number of constellation points inside each coarse lattice cell and rate $R$ denotes the number of bits per complex symbol for $d_k$.

\textbf{Cosets:} Each data symbol $d_k$ now corresponds to a coset of $\Lambda_f$ where
\vspace{-1ex}
\begin{equation}
\label{eq:coset_definition}
\vspace{-1ex}
v_k + \Lambda_c, v_k \in \Lambda_f \cap \mathcal{V}(\Lambda_c),
\end{equation}
with $\mathcal{V}(\Lambda_c)$ denoting the Voronoi region of the coarse lattice $\Lambda_c$, which is a square in the complex plane (of area $\Delta^2$), with distortion bounded by $\Delta / \sqrt{2}$ in Euclidean norm for nearest-neighbor quantization.

Note that the transmit constellation is the set of all coset representatives inside a single Voronoi cell, and not the entire coset.
For a more compresensive overview of nested lattice codes, the reader is referred to \cite{Zamir_Nazer_Kochman_Bistritz_2014}.

\vspace{-2ex}
\subsection{Transmitter Design}

Reintroducing the time-varying component for the data, let each transmit user know its own encoded data symbol $v_k^{(t)} \in \Lambda_f \cap \mathcal{V}(\Lambda_c)$ and computing symbol picked from the rotated set $\mathcal{S}_{\text{rot}} \triangleq \mathbf{R} \, \mathcal{S} = \bigl\{ \mathbf{R} \, s_k \,\big|\, s_k \in \Lambda_f \cap \mathcal{V}(\Lambda_c) \bigr\}$ to maximize orthogonality, where
\vspace{-1ex}
\begin{equation}
\vspace{-1ex}
\mathbf{R} \triangleq \frac{1}{\sqrt{2}}\begin{bmatrix} 1 & -1 \\ 1 & 1 \end{bmatrix}.
\end{equation}

Nested lattice-based \ac{DPC} encodes the communications symbols $v_k^{(t)}$ leveraging a modulo operation such that the transmit symbol $\tilde{v}_k^{(t)}$ is given by
\begin{equation}
\label{eq:lattice_DPC_encoding}
\tilde{v}_k^{(t)} = \text{modL}(v_k^{(t)} - s_k, \Delta),
\end{equation}
where the $\text{modL}(a,b)$ operation is defined as\footnote{An optional dither $\tilde{u}_k$ can be considered, uniformly distributed over $\mathcal{V}(\Lambda_c)$ and known to both the \ac{TX} and \ac{RX} to randomize transmission, but is ignored for simplicity.}
\begin{eqnarray}
\text{modL}(a,b) \triangleq &&\\
&&\hspace{-10ex}
\Big(\text{mod}\big(\Real{a}\!+\!\tfrac{b}{2}, b\big)\!-\!\tfrac{b}{2}\Big)\!+\!j \Big( \text{mod}\big(\Imag{a}\!+\!\tfrac{b}{2}, b\big)\!-\!\tfrac{b}{2} \Big),\nonumber
\end{eqnarray}
with $\text{mod}(a,b)$ denoting the standard modulo operation.

Then, the multifunctional transmit signal $x_k^{(t)}$ across $T$ time slots (analogous to the \ac{SotA} expression in equation \eqref{eq:transmit_sig_decomposition}) can be expressed as
\begin{equation}
\label{eq:lattice_DPC_transmit_signal}
x_k^{(t)} = \tilde{v}_k^{(t)} + s_k,
\end{equation}
where we emphasize that the computing symbol $s_k, \forall k$ remains unchanged across $T$ time slots.

\subsection{Receiver Design}

After transmission of $T$ slots over a typical Rayleigh fading channel, the received signal in matrix form can be expressed as
\vspace{-1ex}
\begin{equation}
\vspace{-1ex}
\bm{Y} = \bm{H}  \bm{X} + \bm{W} = \bm{H} (\tilde{\bm{V}} + \bm{S}) \in \mathbb{C}^{N \times T},
\label{eq:received_signal_lattice}
\end{equation}
where similarly to previous definitions, $\tilde{\bm{V}} \triangleq [\tilde{\bm{v}}^{(1)},\dots, \tilde{\bm{v}}^{(T)}] \in \mathbb{C}^{K\times T}$ with $\tilde{\bm{v}}^{(t)} \triangleq [\tilde{v}_1^{(t)},\dots, \tilde{v}_K^{(t)}]\trans \in \mathbb{C}^{K\times1}$ and $\bm{S} = [\bm{s},\dots, \bm{s}] \in \mathcal{S}_{\text{rot}}^{K \times T} \subset \mathbb{C}^{K\times T}$.

Note that different time slots contain different data symbols $\tilde{v}_k$, retaining the full rate in terms of data transmission identical to typical communications systems, while the computing symbols $s_k$ remain unchanged across $T$ time slots to improve the \ac{MSE} performance of the function evaluation.
One can then apply a \ac{LMMSE} estimator per $t$-th time frame at the receiver to obtain $\hat{\bm{x}}^{(t)}$, which can be expressed as
\vspace{-1ex}
\begin{equation}
\vspace{-1ex}
\hat{\bm{x}}^{(t)} = (\bm{H}\herm \bm{H} + \sigma_w^2 \mathbf{I}_K)^{-1} \bm{H}\herm \bm{y}^{(t)}.
\end{equation}     

Next, a temporary variable $\hat{\bm{z}}$ can be calculated via the modulo operation
\begin{equation}
\hat{\bm{z}}^{(t)} = \text{modL}(\hat{\bm{x}}^{(t)}, \Delta),
\end{equation}
which can then be leveraged to estimate $\hat{\bm{v}}^{(t)}$ via
\begin{equation}
\hat{\bm{v}}^{(t)} = \text{modL}\big( \mathcal{Q}_{\Lambda_f}(\hat{\bm{z}}^{(t)}, \delta), \Delta\big),
\end{equation}
where $\mathcal{Q}_{\Lambda_f}(\cdot)$ denotes the nearest-neighbor quantizer to the fine lattice $\Lambda_f$, and is defined as
\begin{equation}
\mathcal{Q}_{\Lambda_f}(a, b)\! \triangleq \! b \Big(\text{round}\big(\tfrac{\Real{a}}{b} - \epsilon\big) + \epsilon \Big) + jb \Big( \text{round}\big(\tfrac{\Imag{a}}{b} - \epsilon\big) + \epsilon \Big),
\end{equation}
with $\text{round}(\cdot)$ denoting the rounding operation to the nearest integer and $\epsilon \triangleq \text{mod}\Big(\frac{\sqrt{M} - 1}{2}, b\Big)$.\newpage

Finally, a hard decision against the constellation $\Lambda_f \cap \mathcal{V}(\Lambda_c)$ can be performed to obtain the final estimate of the encoded data symbols $\hat{\bm{v}}^{(t)}$.

In order to recover the computing symbols, a novel best-case bound\footnotemark\,  can be derived in the form of a maximum likelihood estimator which can be expressed as
\vspace{-1ex}
\begin{equation}
\label{eq:OTAC_recever}
\hat{\bm{s}}\!=\!\arg\min_{\bm{s} \in \mathcal{S}_{\text{rot}}^{K \times 1}} 
\sum_{t=1}^T \big\| \bm{y}^{(t)}\!-\!\bm{H}\big( \mathrm{modL}(\hat{\bm{v}}^{(t)}\!-\!\bm{s}, \Delta)\!+\!\bm{s} \big) \big\|_2^2,
\end{equation}
where each element $\hat{s}_k$ of  $\hat{\bm{s}}$ is chosen from the discrete set $\mathcal{S}_{\text{rot}}$ that has the lowest error across all time instances $T$, naturally improving the \ac{MSE} performance as $T$ increases.

The computational complexity of equation \eqref{eq:OTAC_recever} is dominated by the number of $K$ users, since an exponential scaling is needed to compare with a constellation of size $M$ such that the dominant complexity is given by $M^K$ in order to obtain a maximum likelihood estimate. 

Finally, the target function can be computed as
\vspace{-1ex}
\begin{equation}
\hat{f}(\bm{s}) = \sum_{k=1}^K \hat{s}_k.
\end{equation}

\vspace{-2ex}
\section{Performance Analysis}

\begin{figure}[H]
\centering
\includegraphics[width=\columnwidth]{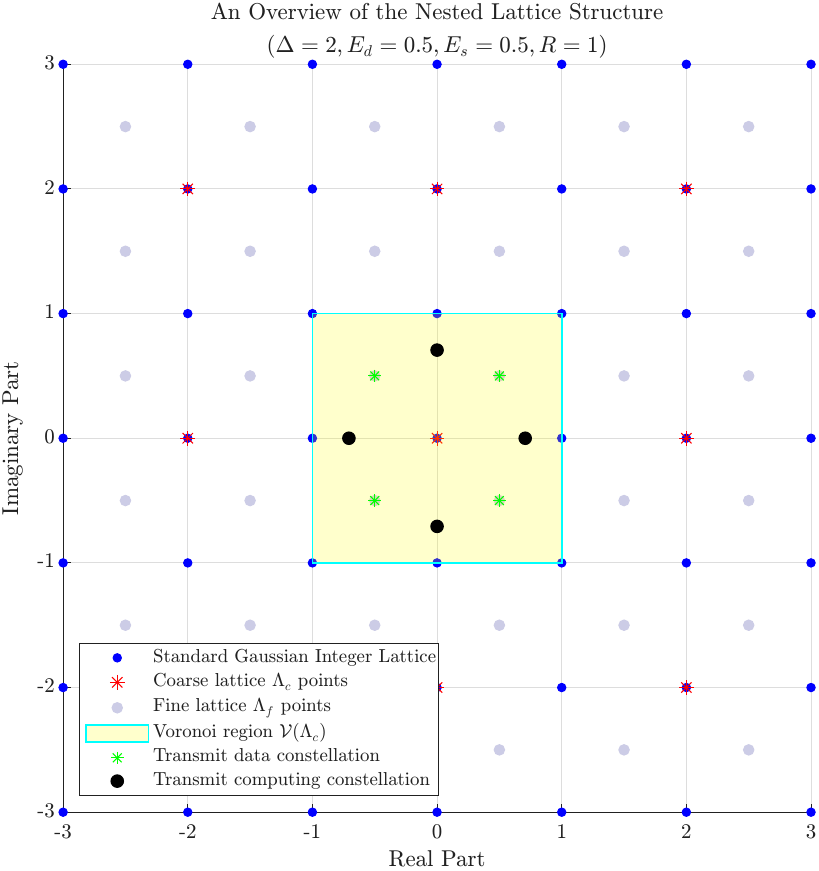}
\vspace{-4ex}
\caption{Nested lattice structure with coarse lattice $\Lambda_c$ and fine lattice $\Lambda_f$.}
\label{fig:lattice_DPC_R1}
\vspace{-2ex}
\end{figure}

To evaluate the performance of the proposed scheme, let us first visualize the complete nested lattice structure that is used, as shown in Fig. \ref{fig:lattice_DPC_R1}.
Notice that each coarse lattice cell contains \( M = 2^{2R} \) constellation points, where \( R \) is the number of bits per complex symbol for data and this fine lattice (corresponding to a scaled \ac{QPSK}) is used for data symbol transmission.
In addition, to maximize orthogonality between data and computing symbols, the computing symbols are chosen from a rotated version of the fine lattice constellation, as previously described.

\vspace{-1ex}

\vspace{-2ex}
\begin{figure}[H]
\subfigure[{\footnotesize $T=5$ (Short Frame Structure)}]%
{\includegraphics[width=\columnwidth]{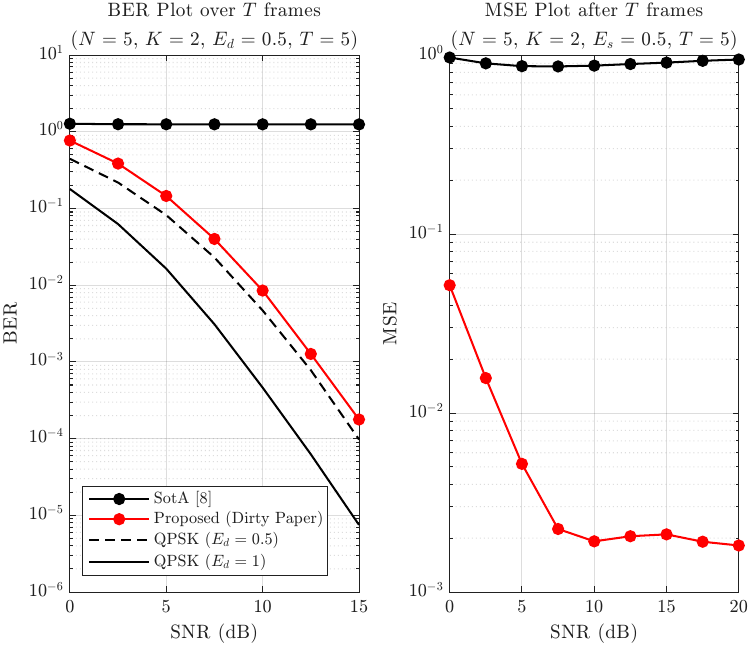}
\label{fig:BER_MSE_T5}}
\vspace{-1ex}\\
\vspace{-1ex}
\subfigure[{\footnotesize $T=10$ (Long Frame Structure)}]%
{\includegraphics[width=\columnwidth]{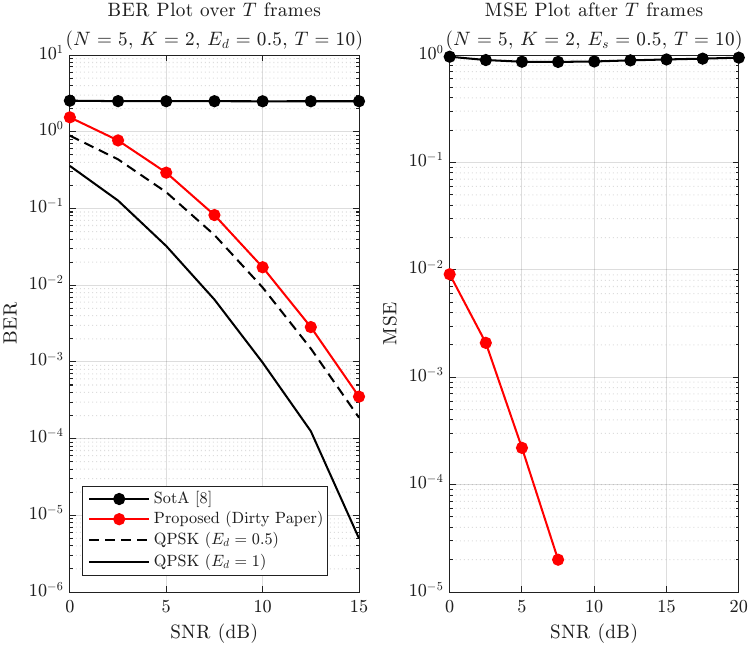}
\label{fig:BER_MSE_T10}}
\vspace{-2ex}
\caption{Comparison of \ac{BER} and \ac{MSE} performances of \ac{SotA} \cite{ranasinghe2025flexibledesignframeworkintegrated} and proposed \textit{computing on dirty paper} \ac{ICC} schemes.}
\label{fig:BER_MSE_Alg2}
\end{figure}

In order to bound the transmit power for a fair comparison with the \ac{SotA}, we set the coarse lattice parameter to be \( \Delta = 2 \), which leads to a transmit power of \( E[|x_k|^2] = E_d + E_s = 0.5 + 0.5 = 1 \) per user, where \( E_d = E[|d_k|^2] = 0.5 \) and \( E_s = E[|s_k|^2] = 0.5 \) are the average powers of data and computing symbols, respectively.

\footnotetext{Although the estimation of each individual computing symbol $s_k$ cannot be considered \ac{OTAC} strictly, this initial estimation procedure provides a proof-of-concept bound for the operation.}

The parameters used in the simulations are \( K = 2 \) users, \( N = 5 \) antennas at the \ac{RX}, an alternating time slot of \( T = 5,10 \), and \( R = 2 \) bits/symbol (i.e., for \ac{QPSK}).

Figure \ref{fig:BER_MSE_T5} and \ref{fig:BER_MSE_T10} show the \ac{BER} and \ac{MSE} performance of the proposed scheme compared to the \ac{SotA} \ac{ICC} method in \cite{ranasinghe2025flexibledesignframeworkintegrated}, where the latter uses a two-stage approach with a \ac{GaBP} detector for data detection followed by a linear combiner for function estimation.
As seen, the proposed scheme achieves significant performance gains in both \ac{BER} and \ac{MSE} compared to the \ac{SotA}, with the performance gap increasing with \ac{SNR}.
This is because the proposed scheme effectively eliminates the interference between data and computing symbols via \ac{DPC}, allowing both tasks to be performed more accurately.
In addition, increasing the number of time slots \( T \) improves the \ac{MSE} performance for the proposed scheme, as more observations are available for function estimation.

\section{Conclusion}
\label{sec:conclusion}

We proposed \textit{Computing on Dirty Paper}, a new scheme for \ac{ICC} inspired by \ac{DPC}.
By pre-canceling computing symbols at the transmitting users, the scheme enables interference-free communication and computation over multiple-access channels. 
Analysis under a \ac{SIMO} \ac{ICC} system and supporting simulations show that the approach achieves superior \ac{BER} and \ac{MSE} performance compared to \ac{SotA} \ac{ICC} methods.

\bibliographystyle{IEEEtran}
\bibliography{references}

\end{document}